\documentclass{ws-ijmpcs}

\begin{document}

\markboth{Zi-Xiang Hu, Ki Hoon Lee and Xin Wan}
{Bulk and edge quasihole tunneling amplitude in the Laughlin state}

%
\catchline{}{}{}{}{}
%

\title{Bulk and edge quasihole tunneling amplitudes in the Laughlin state}

\author{Zi-Xiang Hu}
\address{Department of Electrical Engineering, Princeton University, 
Princeton, New Jersey 08544, USA}
\author{Ki Hoon Lee}
\address{Asia Pacific Center for Theoretical Physics,
Pohang and Department of Physics, Pohang University of Science and
Technology, Pohang, Gyeongbuk 790-784, Korea}
\author{Xin Wan}
\address{Zhejiang Institute of Modern Physics, Zhejiang
University, Hangzhou 310027, P. R. China}

\maketitle

\begin{history}
\received{Day Month Year}
\revised{Day Month Year}
\end{history}

\begin{abstract}
The tunneling between the Laughlin state and its quasihole excitations are studied by using the Jack polynomial. 
We find a universal analytical formula for the tunneling amplitude, which can describe both bulk and edge quasihole 
excitations. The asymptotic behavior of the tunneling 
amplitude reveals the difference and the crossover between
bulk and edge states. The effects of the realistic coulomb interaction with a background-charge confinement potential 
and disorder are also discussed. The stability of the 
tunneling amplitude manifests the topological nature
of fractional quantum Hall liquids.
\keywords{Laughlin state, quasiparticle tunneling, 
Jack polynomial, scaling, disorder}
\end{abstract}  

\ccode{PACS numbers: 73.43.Cd, 73.43.Jn}

\section{Introduction}

Fractional quantum Hall (FQH) liquids are examples of 
experimentally realizable phases that support topological objects. Quasiparticle excitations in the FQH liquids 
have fractional charge and obeys fractional statistics.\cite{Wilczek,Halperin,Arovas} 
Some FQH liquids may support more exotic quasiparticle 
excitations with non-Abelian statistics, which have potential applications in topological quantum computation.\cite{Kitaev}
The measurement of the transport properties of the 
quasiparticles propagating along the edge of FQH liquids
is crucial for the identification of the topological 
nature of the systems. As standard practice in noise 
and interference experiments quantum point contacts are 
introduced to allow quasiparticles propagating on one edge 
to tunnel to another. This motivated the authors and their 
collaborators to study the quasiparticle tunneling amplitudes
in FQH liquids in the disk geometry.\cite{hchen,zxhuNJP} 
We found that the tunneling amplitudes exhibit interesting scaling behavior, whose exponent is related to the conformal
dimension of the tunneling quasiparticles. 

In the disk geometry with an open boundary edge excitations 
arise from the bosonic density deformation of the FQH liquids 
and, in the case of the Moore-Read state, from an extra branch 
of Majorana fermion mode. The edge excitations are closely 
related to the bulk quasihole excitations. For example, a 
charge $\vert e \vert/m$ quasihole at $\xi$ in a $\nu = 1/m$ 
Laughlin droplet 
$\Psi_L = \prod_{i<j} (z_i - z_j)^m 
\exp (-\sum_i \vert z_i \vert^2/4)$ 
of $N$ electrons can be described by the wavefunction
\begin{equation}
\Psi_{qh} = \left [ \prod_i (z_i - \xi) \right ] \Psi_L,
\end{equation}
where $z_j = x_j + i y_j$ is the complex coordinate for the 
$j$th electron. 
The quasihole excitation can be expanded into a sum of 
edge excitations, whose amplitudes depend on the location 
of the quasihole $\xi$
\begin{equation}
\Psi_{qh} = \left [\sum_n (-\xi)^{N-n} s_n \right ] \Psi_L,
\end{equation}
where $s_n = {\cal S}_N (\prod_i^n z_i)$ 
is a symmetric polynomial of degree $n$ and ${\cal S}_N$
denotes the total symmetrization among the $N$ coordinates. 
The first few examples are $s_0 = 1$, $s_1 = \sum_i z_i$, 
$s_2 = \sum_{i<j} z_i z_j$, etc.  
In a Laughlin state the gapless edge mode ${s_n \Psi_L}$ 
spans the Hilbert space of low-energy edge excitations.
In fact, there is no strict distinction between a quasihole 
and an edge excitation with a large angular momentum 
$\Delta M$. 
The conventional understanding is that an edge excitation 
has $\Delta M = O(1)$, while a quasihole excitation 
$\Delta M = O(N)$. 
The correspondence of the bulk and edge excitations suggests
that the bulk and edge quasihole tunneling amplitudes may 
have a parallel correspondence and, therefore, a smooth 
crossover. 

In this paper we confirm that the quasihole tunneling 
amplitude between the Laughlin state and its bulk and 
edge excitations ($\langle \Psi_{qh} \vert {\cal T} \vert 
\Psi_L \rangle$ and $\langle s_n \Psi_L \vert {\cal T} \vert 
\Psi_L \rangle$, where ${\cal T}$ is the tunneling operator
to be defined below), can be described by a unified picture. 
In particular, we conjecture a universal formula in the limit 
of a small interedge distance, which can be reduced to 
the bulk quasihole tunneling amplitude result reported 
earlier.\cite{zxhuNJP} The tunneling amplitude of 
topological nature is robust against the influence of 
long-range interaction and disorder. The paper is organized 
as follows. In Sec.~\ref{review}, we review our previous 
study about the bulk quasihole tunneling amplitude in 
Laughlin state and explain the technical details. 
The study is extended to edge quasihole tunneling 
amplitude in Sec.~\ref{edge}. The robustness of the 
tunneling amplitude in the presence of long-range coulomb interaction and disorder is emphasized in Sec.~\ref{robust}. 
We summarize the results in Sec.~\ref{conclusion}.

\section{A brief review of the bulk quasihole tunneling 
amplitude in a Laughlin droplet}\label{review}

We consider a FQH liquid on a disk and assume a 
single-particle potential $V_{\rm tunnel}(\theta) = V_t \delta(\theta)$, which 
breaks the rotational symmetry. The potential defines a 
tunneling path for quasiparticles under the gate influence 
at a quantum point contact. We start by quoting the 
result\cite{hchen} for the tunneling matrix element 
between two single-particle states 
with angular momentum $k$ and $l$, $v_p (k, l) 
\equiv \langle k| V_{tunnel}(\theta) |l\rangle = \frac{V_t}
{2\pi} \frac{\Gamma((k+l)/2 + 1)}{\sqrt{k!l!}}$. The tunneling operator of the many-body wavefunction is then defined as 
${\cal T} = \sum_i V_{\rm tunnel}(\theta_i)$ and
the amplitude for a quasihole to tunnel to the droplet edge is 
\begin{equation}
\Gamma_{qh} = \frac{\langle \Psi_{qh} \vert {\cal T} \vert 
\Psi_0 \rangle}{\sqrt{\langle \Psi_{qh}\vert 
\Psi_{qh} \rangle}\sqrt{\langle \Psi_0 \vert 
\Psi_0 \rangle}},
\end{equation}
where $|\Psi_0\rangle$ and $|\Psi_{qh}\rangle$ are the 
wavefunctions for the ground state and the quasihole state, 
respectively. 

The matrix element form of the tunneling amplitude suggests
that we can either use Lanczos-type exact diagonalization 
or variational Monte Carlo simulation to calculate. 
However, these approximations fail to reach the accuracy 
needed for the error-free determination of the conformal 
dimensions of quasiholes (though they are useful in the 
discussion of the long-range interaction and disorder 
effects). Fortunately, the application of Jack polynomial 
provides an instructive yet numerical exact calculation 
method.

Let us digress and explain first the connection between 
the Laughlin model wavefunction, which is the exact ground 
state of the hard-core ($V_1$ only) interaction, and the 
Jack polynomial.\cite{Bernevig} 
In general, Jacks belong to a family of {\it symmetric} 
multivariate polynomials of the complex particle coordinates. Potentially, they can be FQH wavefunctions for bosons (appending  the ubiquitous Gaussian factor) or for fermions (with an extra antisymmetric factor $\prod_{i<j} (z_i - z_j)$, i.e., the Vandermonde determinant). 
A Jack $J_\lambda^\alpha(z_1, z_2, \cdots, z_{N})$ 
can be parametrized by a rational number $\alpha$ (negative in this context), which is related to the clustering properties of the polynomial wavefunction, 
and a root configuration $\lambda$, which satisfies a generalized Pauli exclusion 
principle and from which one can derive a set of monomials 
that form a basis for the Jack. The Jack is an eigenstate of 
the corresponding Calogero-Sutherland Hamiltonian
\begin{equation}
H_{CS} = \sum_i (z_i\partial_i)^2 +
\frac{1}{\alpha} \sum_{i<j} \frac{z_i+z_j}{z_i-z_j} (z_i\partial_i - z_j\partial_j),
\end{equation} 
where $\partial_i \equiv \partial /\partial z_i$.
Take the bosonic Laughlin state at $\nu = 1/2$ (which corresponds to the fermionic Laughlin state at $\nu = 1/3$) 
for a concrete example.
One can easily check that $\prod_{i<j} (z_i - z_j)^2$ satisfies
$H_{CS}$ with $\alpha = -2$, which is related to the fact that the bosonic (or the corresponding fermionic) wavefunction vanishes as $(z_i - z_i)^2$ [or $(z_i - z_j)^3$] when particle $i$ approaches $j$. 
For two bosons, one obviously has 
\begin{equation}
(z_1 - z_2)^2 = z_1^2 z_2^0 - 2 z_1^1 z_2^1 + z_1^0 z_2^2 
= 1 \cdot {\cal S}_2 (z_1^2 z_2^0) + (-2) \cdot {\cal S}_2 (z_1^1 z_2^1),
\end{equation} 
which is an expansion of the polynomial wavefunction into 
a sum of symmetric monomials. 
The $N$-particle wavefunction 
$\prod_{i<j} (z_i - z_j)^2$ can be expanded as
\begin{equation}
\label{eq:expansion}
1 \cdot {\cal S}_N (z_1^{2N-2} z_2^{2N-4} \cdots z_N^{0} )
+ (-2) \cdot {\cal S}_N (z_1^{2N-3} z_2^{2N-3} z_3^{2N-6} z_4^{2N-8} \cdots z_N^{0}) 
+ \cdots.
\end{equation}
This crude example (perhaps with a more elaborate expansion) illustrates the idea that a bosonic (fermionic) wavefunction can be expanded by a set of homogeneous symmetric monomials (Slater determinants), which can be derived from a single monomial, known as the root, using a squeezing rule, which lowers the relative angular momentum and conserves the total angular momentum for the system with rotational invariance. 
We choose a numeric string 
representation for the root configuration, which is simply the 
collection of occupation numbers of the single-particle orbitals ($z^m$ in the quantum Hall context). 
The root configuration for the above example of $N$-particle bosonic Laughlin state is, therefore, $1010 \cdots 101$ and for the corresponding fermioic one $100100 \cdots 1001$.
Let us use the convention that the leftmost digit corresponds 
to the $z^0$ orbital, i.e., the droplet center.
It is closely related to the topological nature of the wavefunction 
that the mere knowledge of a Jack parameter 
and a matching root configuration are enough to generate 
the coefficients of all the descendant symmetric monomials (Slater determinant) 
numerically {\it exact} -- practically to more than 10 particles -- with 
a recursive method.\cite{thomale11}

The FQH model wavefunctions can also be written as the correlators of certain primary fields in some conformal field theories. For example, the Laughlin wavefunction at filling fraction $\nu = 1/M$
can be constructed by the chiral boson conformal field theory (CFT) with a compactification radius $M$.\cite{moore91} The primary fields are vertex operators
$e^{i m \varphi(z) / \sqrt{M}}$, where $\varphi(z)$ is a chiral
boson field. Operators with $m = 1, 2, \dots M-1$ correspond to quasiholes with different charge, whose corresponding 
conformal dimensions are $\Delta = m^2/(2M)$.
It is, therefore, reasonable to expect that the tunneling amplitude $\Gamma$ as a function of system size $N$ (with a given tunneling distance $d$) may show scaling behavior, 
$\Gamma \sim N^{\beta}$, whose scaling exponent $\beta$ is related to the conformal dimension of the tunneling particle. 
Our previous work\cite{zxhuNJP} confirmed the hidden (due to 
the dominant single-particle effect) scaling behavior of the
quasihole tunneling amplitude, $\beta = 1 - 2 \Delta$, 
which can be explained by the 
effective field theory consideration that the tunneling amplitudes contains factors of quasiparticles propagating 
along the opposite edges. 

In the 
identification of the relation between the scaling exponent 
and the conformal dimension of the corresponding 
quasiparticle, the exact calculation of the tunneling amplitude 
$\Gamma = \langle\Psi_{qh}|{\cal T}|\Psi_{L}\rangle / (\sqrt{\langle \Psi_{qh}\vert 
\Psi_{qh} \rangle}\sqrt{\langle \Psi_L \vert 
\Psi_L \rangle})$ 
using the 
Jack polynomial method plays a crucial role, which, in fact, motivated the effective 
field theory interpretation.\cite{zxhuNJP} The exactness 
allowed the elimination of all finite-size uncertainties with the conjecture of an exact tunneling amplitude at the small interedge limit, at which we deform the $N$-particle system (with a puncture, or quasiholes, at the center) into a ribbon, or topologically $S^1 \times [0,d]$ ($d/l_B \ll N$). This effectively erases the dominant single-particle 
effect (i.e. the Gaussian Landau level form factor). 
In this limit, for $M = 3$ or $\nu = 1/3$, we conjectured\cite{zxhuNJP} that the tunneling amplitude for
the charge-$e/3$ quasihole is
\begin{equation}
\label{eq:laughlinScaling}
2 \pi \Gamma^{e/M}_{L, M} (N) = {N \over M}
B \left (N, {1 \over M} \right ),
\end{equation}
where $M = 3$ and $N$ is the number of electrons. Here we introduce
the beta function $B(x, \eta) = \Gamma(x) \Gamma(\eta) /
\Gamma(x+\eta)$ which, for large $x$ and fixed $\eta$, asymptotically 
approaches $\Gamma(\eta) x^{- \eta}$, where $\Gamma(x)$ is the Gamma
function (not the tunneling amplitude elsewhere). We verified
numerically that the conjecture is {\it exact for up to 10 electrons};
therefore, asserting the conjecture is also exact for larger system, we
obtain the exact exponent $\beta^{(e/3)} = 1 - 1/3 = 2/3$ in the scaling behavior $\Gamma^{(e/3)} \sim N^{\beta^{(e/3)}}$.
This is also verified to be applicable for $M =
5$.\footnote{Eq.~(\ref{eq:laughlinScaling}) also applies to the
  integer case ($M = 1$), in which the righthand side reduces to
  unity.} In other words, based on the scaling analysis we discussed
earlier, we can compute the conformal dimension of smallest charged quasiholes in the $\nu = 1/M$ Laughlin state to be $\Delta = 1 / (2M)$.

The quasi-one-dimensional ribbon limit essentially removes the unnecessary geometrical information of the wavefunctions, 
and subsequently reveals a perfect scaling behavior 
otherwise embedded 
in inaccuracy and deviations due to small system size. 
A similar consideration, dubbed as the conformal limit, allows 
the opening of a full gap in the entanglement spectra of 
systems on sphere geometry,\cite{thomale10} in a way that
the low-lying levels showing the universal conformal
field theory counting are well separated from 
the higher Coulomb ones, which are not universal. 
In both cases, we found that topology stands out after we suppress the geometrical information. In other words, the topological information is encoded in the set of 
coefficients in Eq.~(\ref{eq:expansion}), while the 
geometrical information is encoded in the monomials, 
which may be deformed, for example, to accommodate 
the geometry of shear transformation and rotation.\cite{read11} 

\section{The edge quasihole tunneling amplitude in a Laughlin droplet}\label{edge}

In the last section, we consider the tunneling amplitude 
of the bulk quasihole at the center of the droplet. In fact, 
we can generalize the discussion to quasiholes located elsewhere, or to edge excitations. 
The generalization is straightforward for the Jack polynomial 
approach, which applies to low-lying excitations, such as the edge mode in disk geometry\cite{kihoon} and the magnetic-roton mode in sphere geometry.\cite{boyang}
The study of the tunneling properties of edge states ($s_n \Psi_L$, $n = 1$, 2, 3, $\dots$) is an important tool for understanding the topological bulk states both from experimental and theoretical point of view. 
Here we focus on the tunneling amplitude 
$\Gamma = \langle s_n \Psi_L|{\cal T}|\Psi_L\rangle$, 
which crossover smoothly from the tunneling of a bulk 
quasihole excitation to edge excitations, 
as $n$ decreases from $N$ to 1. 

The edge mode for the Laughlin state corresponds to a set of states whose root configuration are: ...10010010$\underline{0}$01, ...10010$\underline{0}$01001, ...10$\underline{0}$01001001, etc., meaning one 0 (or a quasihole) can be inserted in any one of the 100 unit cells. The tunneling problem we study here is the amplitude for the quasihole in these edge states to tunnel to the outer edge, leaving the Laughlin state behind. Again, we focus on the quasi-one-dimensional ribbon limit to look for a unified analytical solution. As we have already demonstrated the exactness of the 
method elsewhere,\cite{zxhuNJP} we focus here on presenting 
the analytical conjecture on the tunneling amplitude, which has been verified to be correct for all accessible system sizes. 

For the consistency with Eq.~(\ref{eq:laughlinScaling}), we define, 
\begin{equation}
T(N) \equiv T(N, 1/M) = \sqrt{{2 \pi \over M} \Gamma^{e/M}_{L, M} (N)} = {1 \over M} \sqrt{N
B \left (N, {1 \over M} \right )} ,
\end{equation}
for $N \ge 0$ and specifically define
$T(0, 1/M) = 1/M$. 
This allows us to unify the edge
and bulk quasihole tunneling amplitudes as
\begin{equation} \label{formula}
2 \pi \Gamma^{e/M}_{L, M} (N, \Delta k) = {T(N) T(\Delta k) \over T(N
  - \Delta k)},
\end{equation}
where the integer $\Delta k$ is the angular momentum of the edge/quasihole excitation (i.e., the number of 1s to the right of the inserted 0). For example, if we consider a system 
with a root configuration 10010010010$\underline{0}$01001 and a Jack parameter $\alpha = -2$, the additional parameters can be read as $N = 6$, $M = 3$, and $\Delta k = 2$. When $\Delta k = N$, it recovers Eq.~(\ref{eq:laughlinScaling}), as the additional 0 is located at the leftmost position. 
Note when $\Delta k = 0$, the tunneling amplitude measures 
the average density, which is $1/M$. 
In the thermodynamic limit, for edge excitations, i.e.,
$\Delta k = O(1)$, we find \begin{equation}
2 \pi \Gamma^{e/M}_{L, M} (N, \Delta k) = T(\Delta k) + O(1/N).
\end{equation} 
For bulk excitations, i.e., $\Delta k = O(N)$, we find 
\begin{equation}
2 \pi \Gamma^{e/M}_{L, M} (N, \Delta k) \sim N^{1 - 1/M}.
\end{equation} 
These results allow us to compute straightforwardly the 
tunneling amplitude for a quasihole anywhere inside 
a Laughlin droplet to the edge. 

\section{Robustness of tunneling amplitude in the presence of
realistic edge confinement and disorder}\label{robust}

\begin{figure} 
\begin{center}
 \includegraphics[width=9cm]{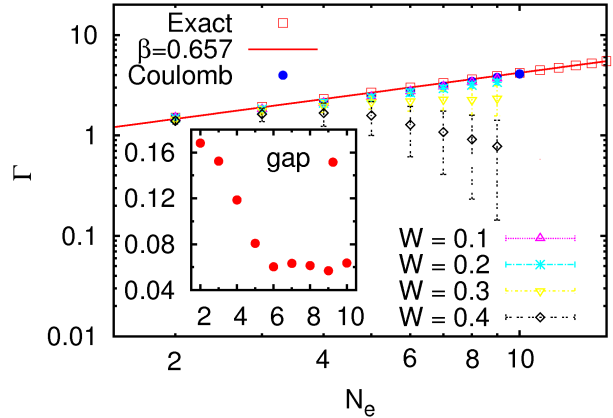}
\caption{\label{fig1} The tunneling amplitude for the Laughlin 
phase in the presence of long-range Coulomb interaction and disorder 
in the quasi-one-dimensional ribbon limit. The exact 
results for the Laughlin state [Eq.~(\ref{eq:laughlinScaling})] 
can be fit by a power law with an exponent 0.657 
(the exact value should be 2/3, which is prone to finite-size 
error in fitting).
In the realistic model with Coulomb interaction and neutralizing 
charge confinement at a setback distance $d=0.5 \ l_B$ (solid points), 
we obtain almost the same $\Gamma$ as using the model wavefunctions. 
When we include the disorder potential with strength $W$, 
$\Gamma$ remains unchanged at weak disorder, 
but deviates from the exact values at strong disorder. 
The inset plots the energy gap between the ground state and 
the first excited state in the same angular momentum subspace with $M=3N(N-1)/2$ 
of the pure Coulomb Hamiltonian with $d=0.5 \ l_B$.}
\end{center}
\end{figure}

So far we discussed the exact tunneling results using model 
wavefunctions generated as Jack polynomials. 
In other words, the Laughlin wavefunction and its edge states (including the single-quasihole state)
are the eigenfunction of the Hamiltonian for 
electrons with hard-core interaction. 
In a realistic GaAs/GaAlAs heterostructure,
the electrons interact with each other via a long-range coulomb repulsion
and are confined by a neutralizing background charge confinement from a doping layer at a setback distance $d$. 
To see whether the tunneling amplitude is robust, we need to verify the validity 
of Eq.~(\ref{eq:laughlinScaling}) in the presence of Coulomb interaction. 
Based on a previous study,\cite{xinprl02} we fix the setback distance of the background charge at $d=0.5 \ l_B$, 
at which the system is in the Laughlin phase, i.e., the global ground state 
has the same quantum number as that of the Laughlin state and 
a close-to-unity overlap with the latter as well. 
On the other hand, the quasihole state is produced by a Gaussian impurity potential 
$H_W = W_g \sum_m \exp(-m^2/2s^2)c^+_m c_m$ with a width $s = 
2 \ l_B$,\cite{zxhuPRB08} which models, e.g., the STM tip potential in an experiment. 
We calculate the bulk quasihole tunneling amplitude up to 10 electrons and compare it (deformed to the quasi-one-dimensional ribbon limit) with Eq.~(\ref{eq:laughlinScaling}).
As shown in Fig.~\ref{fig1},
the long-range coulomb interaction has very little effect on the tunneling amplitude, 
hence the scaling behavior of Eq.~(\ref{eq:laughlinScaling}) is, to a good approximation, valid 
for the realistic interaction, as long as the system remains in the Laughlin phase.

In addition, a realistic system also contains impurity scatterings. 
Nevertheless, the topological properties of an FQH state is believed to 
be robust against weak disorder. 
To prove the statement regarding disorder, we consider 
an uncorrelated random potential on each Landau level orbital, 
such that $H_D = \sum_m U_m c_m^+c_m$, 
where $U_m$ denotes the random potential on the $m$th orbital, 
whose value is randomly chosen in the range of $[-W/2, W/2]$. 
We compute the tunneling amplitudes by averaging over more than 
1000 random samples for a given disorder strength $W$. 
As shown in Fig.~\ref{fig1},  
the tunneling amplitudes for weak disorder, i.e., $W = 0.1$, 
are almost the same as that in the pure coulomb case 
for all accessible system sizes. 
However, when we increase the strength of disorder gradually, 
the tunneling amplitude $\Gamma$ deviates significantly 
from the exact results for large enough system size and 
therefore, the scaling hypothesis of Eq.~(\ref{eq:laughlinScaling}) 
fails in the strong disorder case.
To quantitatively understand the disorder effect, 
we can define and compute the energy gap for the system 
as the energy difference between 
the ground state (which is the Laughlin-like state) 
and the first excited state in the same subspace 
with a total angular momentum $3N(N-1)/2$. 
As shown in the inset of Fig.~\ref{fig1}, when we fix $d = 0.5 \ l_B$, 
the energy gap has very little finite size fluctuations for $N>6$; 
the gap is around 0.06 $e^2/\epsilon l_B$. 
Therefore, disorder starts to affect the tunneling amplitudes 
when the strength of disorder is significantly larger than 
the energy gap (up to an $O(1)$ prefactor). 
This, in return, suggests that the deviation of the tunneling amplitude 
from the scaling behavior can be explored to study the transition 
of the FQH phase to insulator, much like the Chern number study 
of the FQH-insulator transition.\cite{sheng03}

\section{Conclusions}\label{conclusion}
In summary, we calculate the tunneling amplitude for quasihole in the 
Laughlin phase, generalizing the previous result for the tunneling of 
a quasihole at the center of a circular Laughlin droplet to an 
arbitrary location. This is achieved by considering the tunneling 
amplitudes between the Laughlin state and its accompanying edge states.
Using the exact Jack polynomial expansion, we showed that 
the bulk and edge quasihole tunneling can be unified by a 
single equation (\ref{formula}) for any system size. 
We also demonstrate that the quasihole tunneling amplitude 
is robust against realistic considerations, such as the long-range 
coulomb interaction, neutralizing background charge confinement, 
and moderate amount of disorder. 

\section*{Acknowledgements}
This work was supported by the National Basic Research Program of China (973 Program) grant No. 2009CB929100, National Natural Science Foundation of China (NSFC) grant No. 11174246, and US DOE grant No. DE-SC0002140. KHL acknowledges the support at the Asia Pacific Center for Theoretical Physics from the Max
Planck Society and the Korean Ministry of Education, Science and Technology.


\end{document}